\documentclass[twoside,twocolumn,9pt]{article}
\usepackage{extsizes}
\usepackage[super,sort&compress,comma]{natbib} 
\usepackage[version=3]{mhchem}
\usepackage[left=1.5cm, right=1.5cm, top=1.785cm, bottom=2.0cm]{geometry}
\usepackage{balance}
\usepackage{times,mathptmx}
\usepackage{sectsty}
\usepackage{graphicx} 
\usepackage{lastpage}
\usepackage[format=plain,justification=justified,singlelinecheck=false,font={stretch=1.125,small,sf},labelfont=bf,labelsep=space]{caption}
\usepackage{float}
\usepackage{fancyhdr}
\usepackage{fnpos}
\usepackage[english]{babel}
\usepackage{array}
\usepackage{droidsans}
\usepackage{charter}
\usepackage[T1]{fontenc}
\usepackage[usenames,dvipsnames]{xcolor}
\usepackage{setspace}
\usepackage[compact]{titlesec}

\usepackage{graphicx}
\usepackage{dcolumn}
\usepackage{bm}
\usepackage{MnSymbol}	
\usepackage{wasysym}%
\usepackage{amsmath} 
\usepackage{color}

\newcommand{\be}{\begin{equation}}
\newcommand{\ee}{\end{equation}}
\newcommand{\bee}{\begin{eqnarray}}
\newcommand{\eee}{\end{eqnarray}}

\usepackage{epstopdf}

\definecolor{cream}{RGB}{222,217,201}

\begin{document}

\pagestyle{fancy}
\thispagestyle{plain}
\fancypagestyle{plain}{

\renewcommand{\headrulewidth}{0pt}
}

\makeFNbottom
\makeatletter
\renewcommand\LARGE{\@setfontsize\LARGE{15pt}{17}}
\renewcommand\Large{\@setfontsize\Large{12pt}{14}}
\renewcommand\large{\@setfontsize\large{10pt}{12}}
\renewcommand\footnotesize{\@setfontsize\footnotesize{7pt}{10}}
\makeatother

\renewcommand{\thefootnote}{\fnsymbol{footnote}}
\renewcommand\footnoterule{\vspace*{1pt}%
\color{cream}\hrule width 3.5in height 0.4pt \color{black}\vspace*{5pt}} 
\setcounter{secnumdepth}{5}

\makeatletter 
\renewcommand\@biblabel[1]{#1}            
\renewcommand\@makefntext[1]%
{\noindent\makebox[0pt][r]{\@thefnmark\,}#1}
\makeatother 
\renewcommand{\figurename}{\small{Fig.}~}
\sectionfont{\sffamily\Large}
\subsectionfont{\normalsize}
\subsubsectionfont{\bf}
\setstretch{1.125} 
\setlength{\skip\footins}{0.8cm}
\setlength{\footnotesep}{0.25cm}
\setlength{\jot}{10pt}
\titlespacing*{\section}{0pt}{4pt}{4pt}
\titlespacing*{\subsection}{0pt}{15pt}{1pt}

\fancyfoot{}
\fancyfoot[RO]{\footnotesize{\sffamily{1--\pageref{LastPage} ~\textbar  \hspace{2pt}\thepage}}}
\fancyfoot[LE]{\footnotesize{\sffamily{\thepage~\textbar\hspace{3.45cm} 1--\pageref{LastPage}}}}
\fancyhead{}
\renewcommand{\headrulewidth}{0pt} 
\renewcommand{\footrulewidth}{0pt}
\setlength{\arrayrulewidth}{1pt}
\setlength{\columnsep}{6.5mm}
\setlength\bibsep{1pt}

\makeatletter 
\newlength{\figrulesep} 
\setlength{\figrulesep}{0.5\textfloatsep} 

\newcommand{\topfigrule}{\vspace*{-1pt}%
\noindent{\color{cream}\rule[-\figrulesep]{\columnwidth}{1.5pt}} }

\newcommand{\botfigrule}{\vspace*{-2pt}%
\noindent{\color{cream}\rule[\figrulesep]{\columnwidth}{1.5pt}} }

\newcommand{\dblfigrule}{\vspace*{-1pt}%
\noindent{\color{cream}\rule[-\figrulesep]{\textwidth}{1.5pt}} }

\makeatother

\twocolumn[
  \begin{@twocolumnfalse}
\vspace{3cm}
\sffamily
\begin{tabular}{m{4.5cm} p{13.5cm} }

& \noindent\LARGE{\textbf{The non-local repercussions of partial jamming in dense granular flows}} \\
\vspace{0.3cm} & \vspace{0.3cm} \\

 & \noindent\large{Prashidha Kharel$^{\ast}$\textit{$^{a}$} and Pierre Rognon$^{\dag}$\textit{$^{a}$}} \\
& \noindent\normalsize{
This paper establishes a link between the non-local behaviour of granular materials and the presence of transient clusters of jammed particles within the flow.
These clusters are first evidenced in simulated dense granular flows subjected to plane shear, and are found
to originate from a mechanism of multiple orthogonal shear banding. 
A continuum non-local model, similar in form to the non-local Cooperative  model, is then derived by considering the spatial redistribution of vorticity induced by these clusters.
The non-locality length scale is thus expressed in terms of the cluster size.
The purely kinematic nature of this derivation indicates that non-local behaviour should be expected in all glassy materials, regardless of their local constitutive law, as long as they partially jam during flow. 
}
\end{tabular}

 \end{@twocolumnfalse} \vspace{0.6cm}

  ]

\renewcommand*\rmdefault{bch}\normalfont\upshape
\rmfamily
\section*{}
\vspace{-1cm}


\footnotetext{\textit{$^{a}$~Particles and Grains Laboratory, School of Civil Engineering, The University of Sydney, Sydney, NSW 2006, Australia.}}
\footnotetext{\textit{$^{\ast}$~Email: prashidha.kharel@sydney.edu.au}}
\footnotetext{\textit{$^{\dag}$~Email: pierre.rognon@sydney.edu.au}}

%



\section{Introduction}
Glassy materials such as foams, emulsions, and granular matter are composed of amorphous assemblies of many particles, be they bubbles, droplets or grains, that interact with their neighbours upon deformation. These interactions underpin rich flow behaviours that are pivotal to a number of application in engineering, geophysics and biophysics.

Different local constitutive laws have been found to predict the flow of glassy materials, including Herschel-Bulkley model for foams and emulsions \cite{sollich_rheological_1998,hohler2005rheology} and visco-plastic model for granular materials \cite{midi2004dense,da2005rheophysics,Jop:2006aa}.  
Dense granular flows can be characterised in terms of two dimensionless numbers, the shear stress ratio $\mu$ and the inertial number $I$, defined as follows,

\be
\mu \equiv \frac{\tau}{P}; \;\;\; I \equiv \dot\gamma \frac{d}{\sqrt{P/\rho}}
\ee

\noindent where $\tau$ is the shear stress, $P$ is the normal stress, $\dot\gamma$ is the shear rate, $d$ is the grain size and $\rho$ is the grain density.
Their local constitutive law can then be expressed in terms of these two dimensionless numbers as,

\begin{subequations}
\begin{align}
\mu(I) &= \mu_0 + bI  & &\text{for } |\mu| > \mu_0\\
I &= 0 & &\text{otherwise}
\end{align}
\label{eq:loc}
\end{subequations}

\noindent where $\mu_0$ is the yield stress ratio and $b$ is a dimensionless parameter close to unity. Although, this model can fully capture the bulk rheology in the plane shear geometry, strong deviations from this model were evidenced in flow geometries involving the proximity of walls \cite{miller2013eddy,rognon2015long} and/or stress gradients \cite{komatsu2001creep,fenistein2003kinematics,reddy_evidence_2011}. These deviations were attributed to non-local behaviours, and a range of non-local models were introduced to capture them \cite{goyon_spatial_2008,bocquet2009kinetic,pouliquen_non-local_2009,miller2013eddy,bouzid2013nonlocal,wandersman2014nonlocal}, four of which are briefly described in the following paragraph.

Firstly, the \textit{Self-Activation} model \cite{pouliquen_slow_2001, pouliquen_non-local_2009} and the \textit{Cooperative} model \cite{goyon_spatial_2008,bocquet2009kinetic} were both developed based on the idea that a plastic event at a given location in the flow induces stress fluctuations propagating nearby that may trigger a shear event at some other position. The former assumes that the amplitude of the stress fluctuations decreases as a function of the distance from the shear with a characteristic length proportional to the grain size. The latter assumes that the extent of stress redistribution is governed by a \emph{cooperativity length}, $\xi$, which itself is a function of the stresses. 
Secondly, the \textit{Gradient} model \cite{bouzid2013nonlocal,bouzid2015non} is based on the gradient expansion of the yield function, $\mathcal{Y} \equiv \mu/\mu_0$, in terms of the inertial number, leading to the following governing equation: $\mathcal{Y} = \mu(I)/\mu_0 \left( 1- \nu (d^2 \nabla^2 I )/I\right )$, where $\nu$ is a phenomenological constant and $\mu(I)$ is the prediction of the local constitutive law (\ref{eq:loc}). This model also predicts a non-local length scale as a function of the stresses. Finally, the \textit{Eddy Viscosity} model \cite{staron2010flow, miller2013eddy, rognon2015long} is inspired from the turbulent flows of Newtonian fluids, in which the increase in apparent viscosity in turbulent flow is attributed to the development of vortices. Hence the effective viscosity, is expressed as a sum of an intrinsic viscosity of the fluid, and an eddy viscosity due to the formation of vortices. The eddy viscosity is estimated using Prandtl's mixing length model and is governed by the size of vortex and their rotating frequency.

Interestingly, the Cooperative model \cite{bocquet2009kinetic} was shown to successfully predict the flow properties of both emulsions \cite{goyon_spatial_2008} and granular materials \cite{kamrin2012nonlocal,henann2013predictive} in many geometries, even though these materials satisfy two different local constitutive laws. This suggests that non-local behaviours may be independent from local behaviours. It also implies that there could exist a common mechanism governing non-locality for both emulsions and granular materials, and possibly for other similar materials.
For dense granular materials, the cooperative model is expressed in terms of a granular fluidity variable, $f$,  defined as the ratio of shear strain-rate to shear stress.

\begin{subequations}
\begin{align}
f &\equiv \dot{\gamma}/\mu \\
\nabla^2 f &= \frac{1}{\xi^2}(f-f_l) \label{eq:nl_pde}\\
\xi &= A \frac{d}{\sqrt{| \mu - \mu_0|}}\label{eq:cooplen}
\end{align}
\label{eq:coop}
\end{subequations}

\noindent Here $f_l$ is the local fluidity predicted by the local constitutive law (\ref{eq:loc}). 

While this continuum model conveniently captures several observed non-local properties of granular flows, its origins in terms of microstructre and micro-mechanical processes remain elusive. Interestingly, the macroscopic fluidity $f$ has recently been shown to be a purely kinematic variable, which can be measured in terms of local  grain velocity fluctuations and solid fraction \cite{zhang2017microscopic}. The fact that fluidity can be associated to grain kinematics leads to two subsequent questions: (i) can its cooperativity length scale $\xi$ be measured from the grain kinematic? And (ii) which kinematic process could lead to the non-local behaviour, as expressed by a partial differential equation such as (\ref{eq:nl_pde})?

The purpose of this paper is to address these two questions. In this aim, we performed a series of DEM simulations of granular flows in various geometries, and analysed the link between their internal kinematic field and their non-local behaviour. The paper is structured as follows. In Sections \ref{sec:ps}, we analyse the kinematic field of homogeneous plane shear flows in absence of walls and stress gradients. We specifically seek to evidence the development of transient kinematic clusters, to measure their size and identify how it scales with the inertial number. In Section \ref{sec:cluster_NL}, we introduce a physical argument which directly connects the existence of these clusters to a non-local continuum model similar to (\ref{eq:nl_pde}), providing us with a tentative answer to questions (i) and (ii). Finally, in Section \ref{sec:test}, we assess the ability of such a cluster-based non-local model to capture the flow profiles in geometries involving walls and stress gradients.


\begin{figure}[!b]
\centering
\includegraphics[width=1\columnwidth]{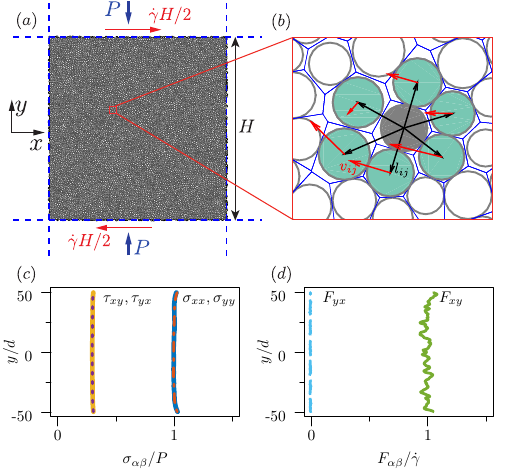}
\caption{\label{fig:system} Homogeneous shear of granular materials. (a) 100$\times$100 grains sheared within a bi-periodic domain; (b) Close up of a grain (grey) and its Voronoi neighbours (green); (c-d) Profiles of stresses and velocity gradients for a system of frictional grains with $I=0.01$ (averaged values over $\gamma = 50$ shear deformation on strips of size $0.5d$).}
\end{figure}

\section{Plane shear flow} \label{sec:ps}
In order to identify what property of the microstructure could govern the non-local length scale, we performed discrete element simulations of dense granular flows in which the motion of each grain, both translation and rotation, is integrated over small time steps using a second-order predictor-corrector scheme, as in\cite*{rognon2015long}. A system of $10\,000$ grains in a 2D periodic domain is subjected to shear, prescribing both the shear rate $\dot \gamma$ and the normal stress $P$ (see Fig. \ref{fig:system}(a)). Grains have a polydispersity of $\pm20\%$ on their diameter $d$ in order to avoid crystallisation. They interact with their neighbours \textit{via} inelastic and frictional contacts, characterised by a Young's modulus of $E = 10^3 P$, a coefficient of restitution $e=0.5$ and a coefficient of friction $\mu_g = 0.5$.
The advantage of the plane shear geometry is to produce homogeneous steady flows in which stresses and shear rate are constant throughout the shear layer (see Fig. \ref{fig:system}(c-d)). The use of Lees-Ewards periodic boundary conditions \cite{lees1972computer} prevents the introduction of walls and avoids the shear heterogeneity they would induce \cite{rognon2015long}. 
Several steady flows of differing shear rate $\dot \gamma$ were performed within the dense regime \cite{midi2004dense}, covering the following range of inertial number: $5 .10^{-4} \leq I \leq 10^{-1}$.

\subsection{Observing kinematic clusters}\label{sec:cluster}

\begin{figure}[!t]
\includegraphics[width=\columnwidth]{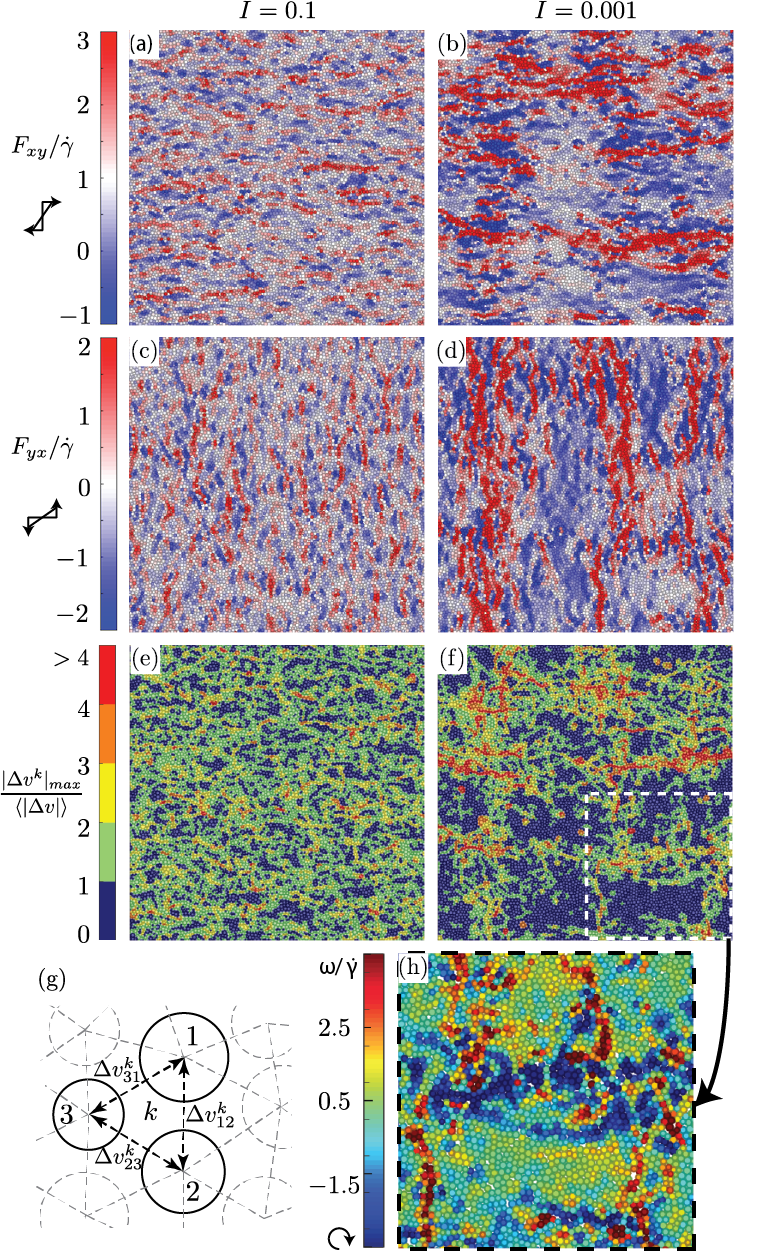}
\caption{
\label{Fig:ShearBands} 
Multiple shear bands and clusters of jammed grains in plane shear flows for two inertial numbers.
(a-d) First two rows are snapshots of spatial distribution of the velocity gradients along $x$ direction and $y$ direction respectively. 
(e-f) Snapshots of maximum relative velocities for each triplet of grains (see text); Partially jammed triplets are shown in dark blue.
(g) Schematic of Delaunay triangulation and computation of relative velocities for triplet of grains. (h) A closeup of (f) showing the local vorticity $\omega = \frac{1}{2}(F_{xy}-F_{yx})$. (Movies available online).
}
\end{figure}

\begin{figure}[!t]
 \centering
\includegraphics[width=\columnwidth]{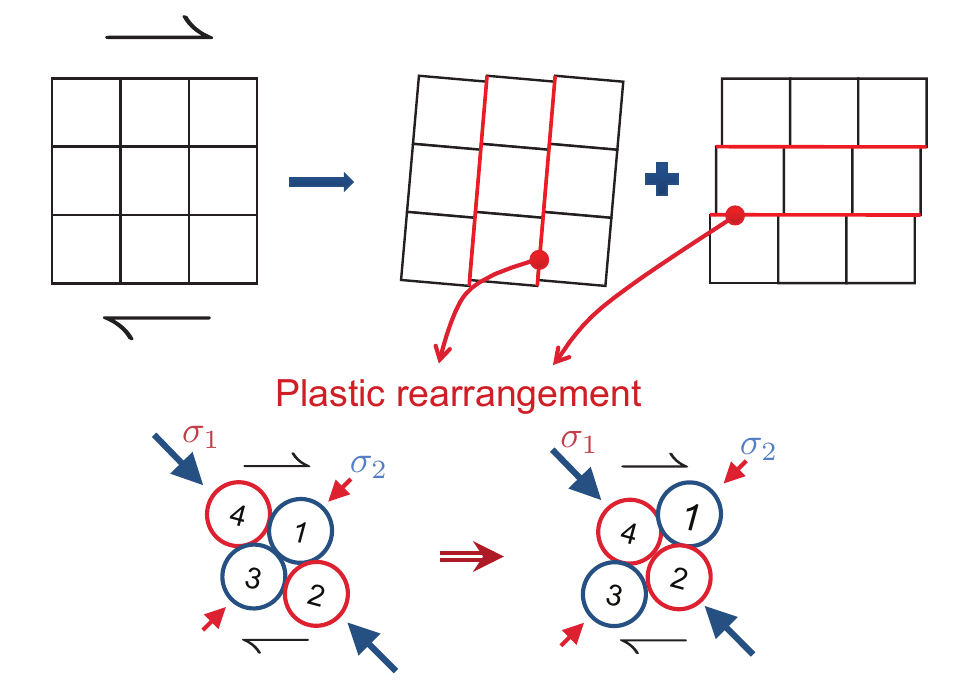}
\caption{
\label{Fig:Plastic} 
Illustration of the mode of deformation involving multiple orthogonal shear bands, where plastic events take place. (Top) Plane shear induce shear bands in two directions; 
(Bottom) Example of an elementary plastic event ($T_1$ process); $\sigma_1$ and $\sigma_2$ are the major and minor principal stresses. 
}
\end{figure}

Figure \ref{Fig:ShearBands}(a-d) show a snapshot of the local velocity gradients within two flows with different inertial numbers. Local velocity gradients are quantified by the tensor $\bm{F}^i$, which is defined for each grain $i$ by considering its relative velocity $\bm{v}^{i,j}$ and distance $\bm{l}^{i,j}$ to its Voronoi neighbours $j$ (as shown in figure \ref{fig:system}(b)) by employing the formula \cite{rognon2015long}:

\begin{equation}
\bm{F}_{\alpha \beta}^i \equiv
\frac{\partial v_\alpha}{\partial x_\beta}\Bigr|_i = 
\langle \bm{l}^{i,j} \otimes \bm{l}^{i,j} \rangle^{-1} \bullet
\langle \bm{l}^{i,j} \otimes \bm{v}^{i,j} \rangle
\end{equation}

\noindent where $\bullet$ and $\otimes$ represent the tensor product and outer product, and $\langle \cdot \rangle$ represents  the average over all $j$ neighbours of grain $i$. The components $F_{yx}$ and $F_{xy}$ shown on figure \ref{Fig:ShearBands}(a-d) denote a rate of shear deformation parallel and orthogonal to the flow direction $x$, respectively. It appears that the shear is localised on multiple shear bands both in the flow direction and in the transverse direction. These two directions correspond to the direction of maximum shear stress in the flow, given that the two normal stresses are equal (see Fig. \ref{fig:system}(c)). This mechanism of multiple and orthogonal shear banding creates a lattice of highly sheared zones \cite{falk1998dynamics, falk2011deformation} delimiting the boundaries of cluster of grains subjected to little if any shear deformation. Figure \ref{Fig:Plastic} illustrates this mechanism. Figure \ref{Fig:ShearBands}(h) shows that the local vorticity is nearly constant within clusters with a value close to $\dot\gamma /2$, which further confirms that clusters rotate like rigid bodies.

\begin{figure}[!t]
\centering
\includegraphics[width=1\columnwidth]{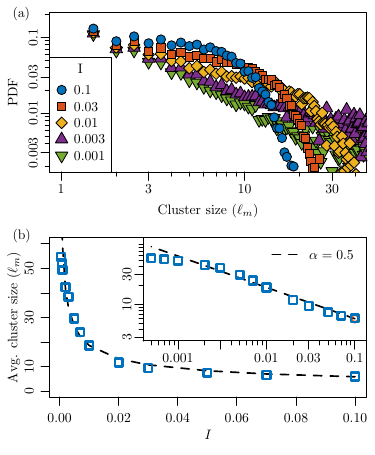}
\caption{\label{fig:pdf_size} Measured cluster size in plane shear flows. (a) Probability distribution of the cluster size for frictional granular flows. 
(b) Average cluster size for frictional granular flows at different inertial numbers. Inset: Same data in log-log scale. The dashed line represent power law fit, $\ell \propto d/I^\alpha$, which is similar to (\ref{eq:xi}). 
}
\end{figure}

\subsection{Measuring cluster size} \label{sec:cluster_size}

As a way to identify individual clusters and measure their size,  we developed the following method. First, triplets of neighbouring grains are identified using a Delaunay triangulation. Then, a criteria is employed to determine which triplets are jammed. Amongst several possibilities, we used a criteria based on the maximum relative velocity between pair of grains in the triplet $k$: $| \Delta v^k|_{max} = max(|\Delta v_{12}^k|;|\Delta v_{13}^k|; |\Delta v_{23}^k|)$ (see Fig. \ref{Fig:ShearBands}(g)). The triplet is considered kinematically jammed if all of the three relative velocities are smaller than the average relative velocity $\langle | \Delta v|\rangle$ between all pair of neighbouring grains in the full system. 
Such  partially jammed triplets are shown in dark blue in figures \ref{Fig:ShearBands}(e-f). Finally, an aggregating algorithm is used to connect adjacent jammed triplets into clusters. The measured size $\ell_m$ of a cluster of $n$ grains is then defined as $\ell_m = d\sqrt{n}$. 

Figure \ref{fig:pdf_size} shows the cluster size distribution obtained by using the above criteria on $1000$ snapshots evenly distributed over a total shear deformation of $50$. Strikingly, the PDF follows a power law with a rate-dependent exponential cutoff, implying that the cluster sizes are not scale-free. The value of this cutoff increases for decreasing inertial number, showing that larger clusters are more likely to appear in systems with smaller inertial numbers. Moreover, the average cluster size $\ell$ scales with the inertial number with a power law, as shown on figure \ref{fig:pdf_size}(b):

\be \label{eq:xi}
\ell \propto \frac{d}{I^{\alpha}}
\ee

\noindent With our frictional grains, our data suggest a power $\alpha=0.5$ as long as the cluster size remains smaller than the system size\cite{PhysRevLett.119.178001}. This power is consistent with other internal length scale measured in granular flow\cite{staron2008correlated,lemaitre2009rate,halsey2005coherent,DeGiuli:2015aa,degiuli2015unified,rognon2015long}. 

A rational for the scaling (\ref{eq:xi}) can be deduced by considering the fact that flow in the dense regime occurs primarily due to localised plastic rearrangements. If we assume that a typical plastic rearrangement, like the T1 process \cite{rognon2009soft} shown in figure \ref{Fig:Plastic}, will produce a net rearrangement comparable to the average particle size $d$ under an average pressure $P$, then such a plastic event will produce 1 shear deformation and will last for some finite relaxation time $t_R  \propto d\sqrt{\rho/P}$, also called the inertial time.  In the dense regime, the inertial time will be very small compared to the shear time $1/\dot{\gamma}$, i.e. $t_R << 1/\dot{\gamma}$ \cite{midi2004dense,da2005rheophysics}. 
But the kinematics of the flow requires 1 strain to be produced in $1/\dot{\gamma}$ time, i.e. the shear time. This means localised plastic events cannot occur everywhere simultaneously and continuously. One option is for them to occur heterogeneously, with some typical separation $\ell$ between simultaneous plastic events in the two orthogonal directions. Accordingly, in average, a sequence of $\ell^2/d^2$ plastic events spanning over an area $\ell^2$ can produce a net strain of 1 over that area and this would last $(\ell^2/d^2)t_R$ time. Equating this time to the shear time $1/\dot\gamma$ required by the kinematics of the flow to produce 1 strain deformation, leads to the following relation for the separation length:

\be \frac{\ell^2}{d^2} t_R = \frac{1}{\dot{\gamma}}; \;\;\;\ell = \frac{d}{\sqrt{\dot{\gamma} t_R}} = \frac{d}{\sqrt{I}}\label{eq:ell_derive}\ee 

\noindent This separation, $\ell$, can also be interpreted as the size of the region where plastic events are not taking place at a given time, i.e., the region where the grains are temporarily jammed.  

Most importantly, the measured scaling of the cluster size (\ref{eq:xi}) is similar to the scaling of the cooperativity length $\xi(I)$ used in continuum  models \cite{kamrin2012nonlocal,bouzid2013nonlocal,ries2017cooperativity}, as in Eq. (\ref{eq:cooplen}), considering that $\mu -\mu_0 \propto I $ for $\mu>\mu_0$. This suggests that jammed clusters could  somehow be related to the non-locality length scale.

\section{Clusters and non-locality} \label{sec:cluster_NL}

We now seek to establish how the existence of jammed clusters can give rise to non-local behaviours at the continuum scale. 
The key consideration is that clusters move as rigid bodies, and therefore redistribute their vorticity over their size (see Fig. \ref{Fig:ShearBands}(h)). 
The local value of the vorticity within a flow should therefore be affected by the value of the vorticity in the surrounding. In the following a non-local continuum model is derived based on this mechanism of vorticity redistribution.

 \begin{figure}[!t]
 \centering
\includegraphics[width=0.9\columnwidth]{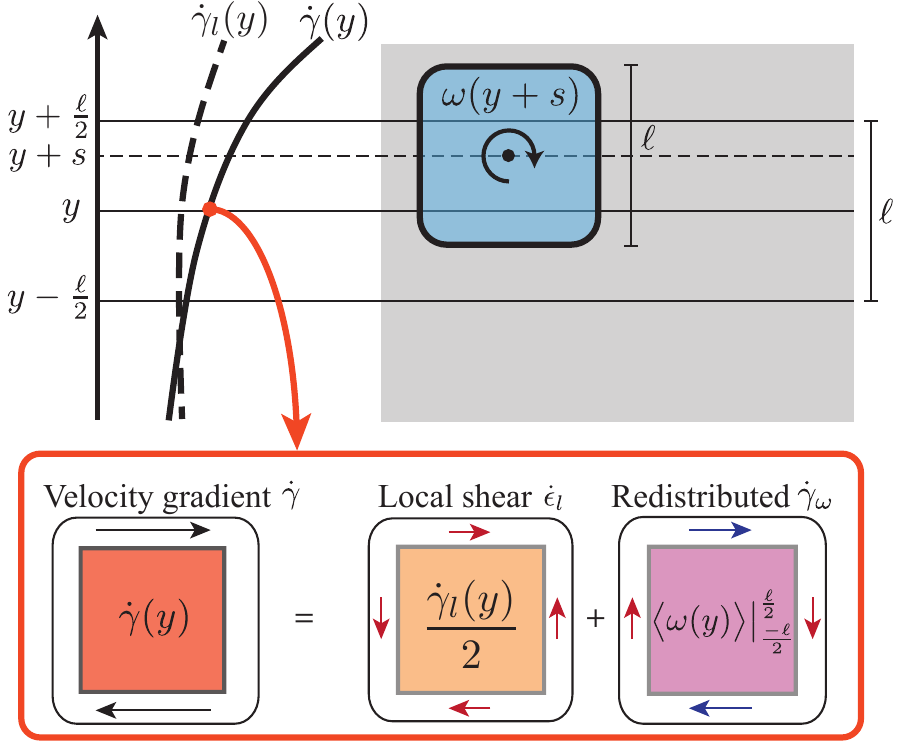}
\caption{\label{fig4} 
Illustration of the proposed mechanism at the origin of non-locallity. \textit{(Top)} Actual shear rate profile $\dot \gamma(y)$ in a heterogeneous sheared layer compared to the shear rate profile $\dot \gamma_l$ predicted by a local constitutive law with no account for non-local effects; A clusters of jammed particles located at a position $y+s$ is represented, distributing its vorticity over its size $\ell$. 
\textit{(bottom)} The local velocity gradient is comprised of i) a local pure shear strain rate $\dot\epsilon = \dot\gamma_l/2$ governed by the local stresses via a local constitutive law, and a contribution $\dot\gamma_\omega = \langle \omega(y)\rangle|_{-\ell/2}^{\ell/2}$ coming from the vorticities of nearby clusters.
}
\end{figure}

Within a homogeneously sheared layer subjected to a bulk shear rate $F_{xy}=\dot \gamma_l$ and $F_{yx}=0$ (see Fig.\ref{fig:system}(d)), the bulk shear rate $\dot\gamma_l$ will depend on the state of stress and it can be back calculated using the local constitutive law of the material, for instance the relation (\ref{eq:loc}) for granular materials \cite{midi2004dense,da2005rheophysics,Jop:2006aa}. This deformation can be decomposed into  a pure shear deformation, $\dot\epsilon_l = \dot\gamma_l/2$, and pure rotation represented by the vorticity, $\omega_l = \dot\gamma_l/2$: $\dot\gamma_l = \dot\epsilon_l + \omega_l$. The shear deformation is responsible for local mechanical dissipation, and the vorticity governs the cluster rotation rate which gets spatially redistributed. 
This decomposition also applies locally in a non-homogeneously sheared layer in which the shear rate depends on $y$: $\dot\gamma(y) = \dot\epsilon(y) + \omega(y)$.
We further decompose the velocity gradient into a local shear strain deformation that is not redistributed, $\dot\epsilon_l$, and a shear rate due to redistribution of vorticity, $\dot\gamma_\omega$ (see Fig. \ref{fig4}):

\be \label{eq:shear_decouple2}
\dot\gamma(y) = \dot\epsilon_l(y) + \dot\gamma_\omega(y)
\ee

\noindent In the homogeneous case, the vorticity is constant throughout all layers and the shear rate due to redistribution of vorticity $\dot\gamma_\omega$ is therefore equal to the vorticity $\omega$ and will contain only the rotational component. By contrast, in the non-homogeneous case, this may not be the case and $\dot\gamma_\omega$ may contain some pure shear component along with rotation depending on the geometry, as illustrated in Fig \ref{fig4}.

When a cluster of size $\ell$ develops at some position $y+s$, it redistributes the vorticity $\omega(y+s)=\dot\gamma(y+s)/2$ over the zone comprised over its size $\ell$. 
Given that these clusters are transient and develop at different locations, the net shear rate at a point $y$ due to redistribution of vorticity is given by the average of the vorticities of points between $y\pm\ell/2$, i.e., 
$\dot\gamma_\omega(y) =  \langle \omega(y) \rangle |_{-\ell/2}^{\ell/2} =\frac{1}{\ell} \int_{-\ell/2}^{\ell/2} (\dot\gamma(y+s)/2) ds $. 
By contrast, jammed clusters do not redistribute local pure shear, which is entirely governed by the local stresses in the layer and is therefore the same as in the case of homogeneous shear flow, i.e.,  $\dot\epsilon_l(y) = \dot\epsilon_l(y)=\dot\gamma_l(y)/2$.
Substituting these two results in (\ref{eq:shear_decouple2}) leads to:

\be \label{eq:init_shear_eq}
2\dot\gamma(y) = \dot\gamma_l(y) + \frac{1}{\ell} \int_{-\ell/2}^{\ell/2} \dot\gamma(y+s) ds
\ee

\noindent By using a Taylor expansion of $\dot\gamma(y+s)$ with respect to $s$ about $s=0$, the integral in (\ref{eq:init_shear_eq}) becomes:

$$\frac{1}{\ell} \int_{-\ell/2}^{\ell/2} \left(
\dot\gamma(y) + 
\frac{\partial \dot\gamma(y)}{\partial y}s +
\frac{1}{2}\frac{\partial^2 \dot\gamma(y)}{\partial y^2}s^2 +
\mathcal{O}\left(\frac{\partial^3 \dot\gamma}{\partial y^3}\right)
\right) ds$$

$$
= \dot\gamma(y) + 
\frac{\ell^2}{24}\frac{\partial^2 \dot\gamma(y)}{\partial y^2} +
\mathcal{O}\left(\frac{\partial^4\dot\gamma}{\partial y^4}\right)
$$

\noindent  Introducing the second order approximation of this expression into (\ref{eq:init_shear_eq}) leads to a non-local equation governing the shear rate $\dot \gamma(y)$: 

\be \label{eq:NL_continuum}
\dot{\gamma}(y) - \dot{\gamma}_{b}(y) = \frac{\ell^2}{24} \frac{\partial^2 \dot{\gamma}}{\partial y^2}.
\ee

\noindent This expression is similar to the non-local Cooperative model (\ref{eq:coop}) which is written in terms of fluidity $f$.
Two different definitions were used for the fluidity, $f = \dot\gamma/\tau$ for emulsions \cite{goyon_spatial_2008} and $f = \dot\gamma/\mu$ for granular materials \cite{kamrin2012nonlocal,henann2013predictive}. 
In both cases, non-local effects arise when flows are near the jamming transition. If we ignore the second order gradient of the shear stress $\tau$ for emulsions and that of shear stress ratio $\mu$ for granular materials, the formulation in terms of fluidity then reduces to (\ref{eq:NL_continuum}) and the cooperativity length is directly given by the average cluster size: 

\be \label{eq:24}
\xi \approx \ell/\sqrt{24}
\ee

\begin{table}[h!]
  \small
  \centering
    \caption{DEM Simulation parameters for the two geometries: Plane Shear with Gravity (PSG) and Poiseuille flow (PF).}
    \label{tab:config}
    \begin{tabular}{c|c|c|c|c|c} 
      Symbol & Geometry & $H/d $&$10^3P_w/E$ & $v_w \sqrt{\rho/P_w}$ & $g\sqrt{\rho/P_w}/d$\\
      \hline
      {\color[rgb]{0, 0.4470, 0.7410}$\blacksquare$} & PSG  &60&0.4 & 0.316 & 0.0095\\ 
      {\color[rgb]{0.8500,    0.3250,    0.0980}$\square$} & PSG  & 60 & 0.4 & 0.791 & 0.0126\\ 
      {\color[rgb]{0.9290,    0.6940,    0.1250}$\blacktriangle$} & PSG  &30 & 0.4 & 0.316 & 0.019\\ 
      {\color[rgb]{0.4940,    0.1840,    0.5560}$\triangle$} & PSG  &30 & 0.4 & 0.791 & 0.019\\
      {\color[rgb]{0, 0.4470, 0.7410}$\bullet$} & PF &80& 1 & - & 0.01\\ 
      {\color[rgb]{0.8500,    0.3250,    0.0980}$\circ$} & PF & 80 & 1 & - & 0.0125\\ 
      {\color[rgb]{0.9290,    0.6940,    0.1250} $\blacktriangledown$} & PF &40 & 1 & - & 0.036\\ 
      {\color[rgb]{0.4940,    0.1840,    0.5560}$\triangledown$} & PF &40 & 1 & - & 0.024\\ 
      
    \end{tabular}
\end{table}

 \begin{figure}[!t]
 \centering
\includegraphics[width=1.05\columnwidth]{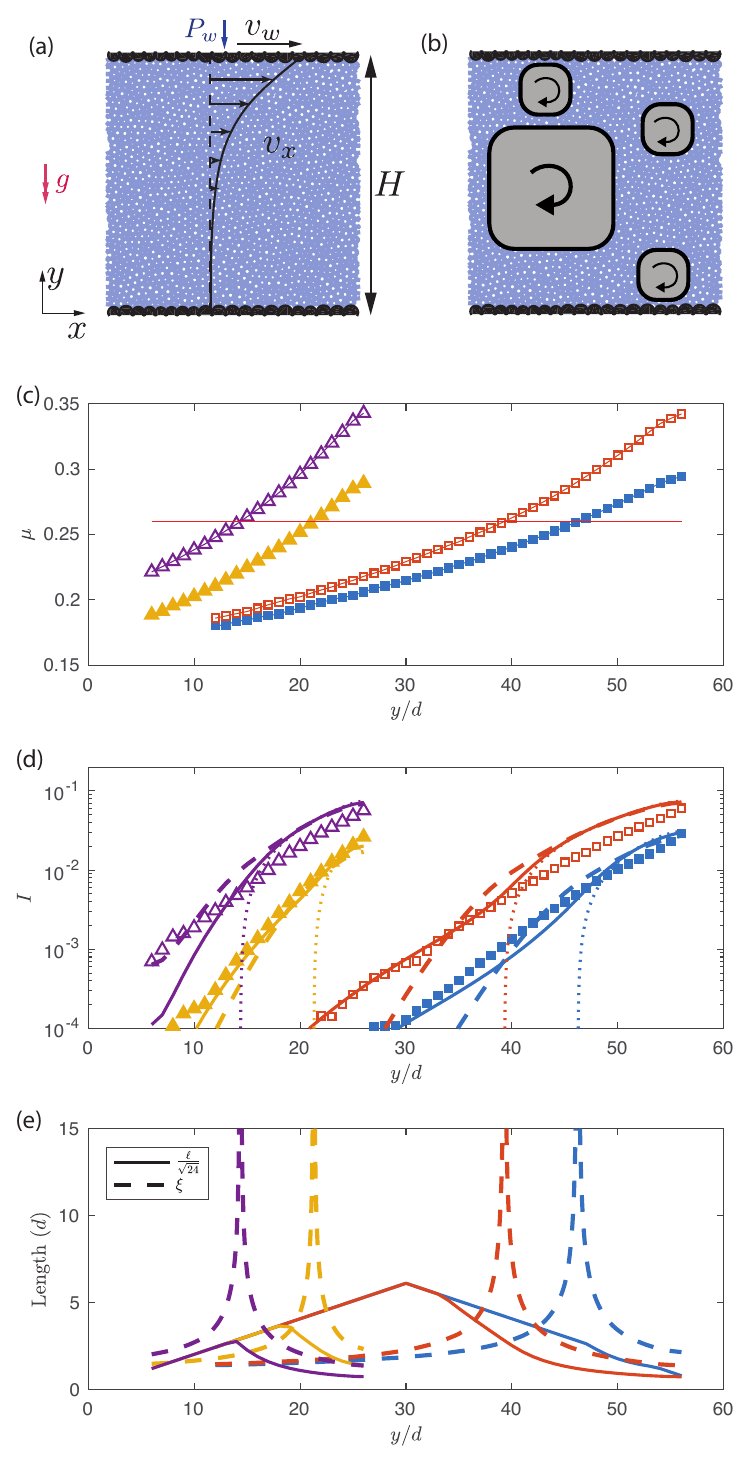}
\caption{\label{fig:sg} 
Plane shear with Gravity (PSG). 
(a) $H\times H$ Plane shear with gravity geometry where wall grains (black) have an average diameter $2d$, with the bottom wall stationary and top wall subjected to a vertical pressure $P_w$ and horizontal velocity $v_w$, whereas the flowing flowing grains (blue) have average diameter $d$ and are subjected to a constant force of $\pi d^2 \rho g/4$ along $y$ axis. 
(b) Illustration of partially jammed clusters of grains within the flow.
(c) Shear stress ratio profile, (d) shear rate profile and (e) length scales as a function of $y$ where the symbols represent data from DEM simulations with parameters given in table \ref{tab:config}, dotted curves are prediction of local constitutive law (\ref{eq:loc}), dashed curves are solutions of cooperative model (\ref{eq:coop}) and solid curves are solutions of the cluster based model in eqs. (\ref{eq:NL_continuum}) and  (\ref{eq:ell_cap}).
}
\end{figure}


 \begin{figure}[!t]
 \centering
\includegraphics[width=1.05\columnwidth]{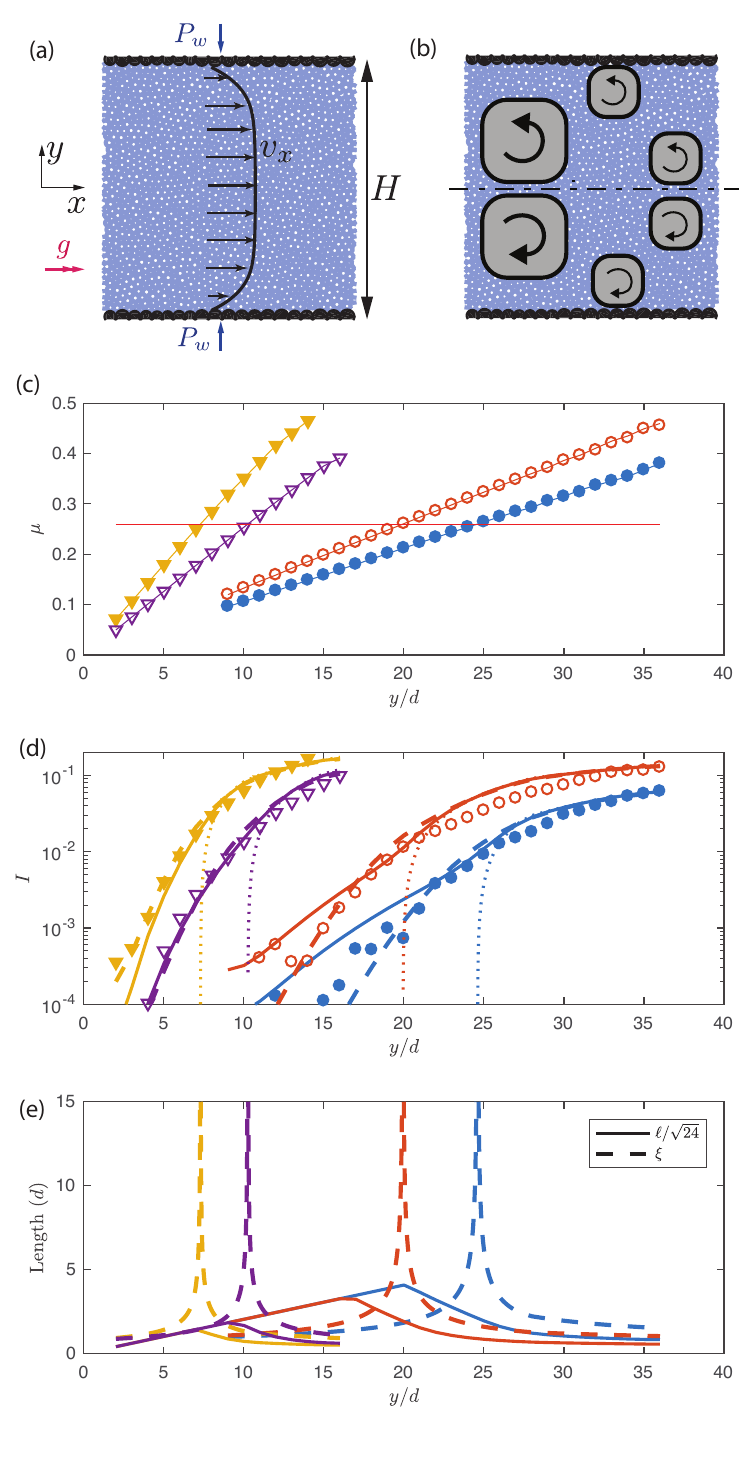}
\caption{\label{fig:vc} 
Poiseuille Flow (PF). 
(a) $H\times H$ Poiseuille flow geometry where wall grains (black) with average diameter $2d$ are stationary along $x$ axis and subjected to a pressure $P_w$ along $y$ axis, whereas flowing grains (blue) with average diameter $d$ are subjected to a constant force of $\pi d^2 \rho g/4$ along $x$ axis. 
(b) Illustration of partially jammed clusters of grains within the flow.
(c) Shear stress ratio profile, (d) shear rate profile and (e) length scales as a function of $y$ where the symbols represent data from DEM simulations with parameters given in table \ref{tab:config}, the dotted curves are prediction of local constitutive law (\ref{eq:loc}), the dashed curves are solutions of cooperative model (\ref{eq:coop}) and solid curves are solutions of eqs. (\ref{eq:NL_continuum}) and (\ref{eq:ell_cap}).
}
\end{figure}

\section{Non-homogeneous shear flow}\label{sec:test}

As a way to assess the validity of the cluster-based non-local model  (\ref{eq:NL_continuum}), let us now consider a series of granular flow DEM simulations in a plane shear with gravity and Poiseuille flow geometries\footnote{ All reported simulation data are recorded once the flow reaches the steady state. The total recording time is set such that the slowest shearing layer reported undergoes at least 5 shear deformations.} (see Fig. \ref{fig:sg}(a) and Fig. \ref{fig:vc}(a)).
Both of these geometries involves a shear stress ratio gradients (see Fig. \ref{fig:sg}(c) and Fig. \ref{fig:vc}(c)), and the presence of walls, which both induce non-local effects \cite{kamrin2012nonlocal, rognon2015long}. The symbols in Fig. \ref{fig:sg}(d) and Fig. \ref{fig:vc}(d) show the  measured profiles of inertial number for various simulation configurations listed in Table \ref{tab:config}. 

Firstly, we implemented the prediction of the local constitutive law (\ref{eq:loc}) with $\mu_0=0.26$ and $b=1.1$. The value of these parameters were inferred from our plane shear simulations and are consistent with the previousely reported results  \cite{da2005rheophysics, kamrin2012nonlocal,bouzid2013nonlocal, tang2018nonlocal}. Fig. \ref{fig:sg}(d) and Fig. \ref{fig:vc}(d)  confirms that the prediction of the local constitutive law (dotted curves) do not fully capture the inertial number profiles, especially in the layers with shear stress ratio below the yield (i.e. $\mu<\mu_0$). 

Secondly, we implemented the Cooperative model  as per (\ref{eq:coop}) by solving the PDE using a finite difference solver (\emph{bvp5c} in MATLAB) and using Neumann boundary conditions at both ends, setting the gradient of fluidity to be null in these two locations, as in \cite{kamrin2012nonlocal,tang2018nonlocal}. Following the literature \cite{tang2018nonlocal}, we used the constant $A=0.41$ to compute  the cooperativity length $\xi$ from the stresses, as per Eq. (\ref{eq:cooplen}). Fig. \ref{fig:sg}(d) and Fig. \ref{fig:vc}(d)  show the prediction of this model (dashed line), which appears to satisfactorily capture the inertial number profile.

Finally, we implemented the cluster-based non-local model expressed in Eq. \ref{eq:NL_continuum}, using the same parameter for the local constitutive law as above. To define the cluster size, $\ell$, we used the scaling  (\ref{eq:xi}) with $\alpha = 0.5$ evidenced with plane shear flows. To be consistent with the vision that the length $\ell$ represents a cluster size, we truncated its values near walls, and near zones in the flow where the shear stress ratio changes sign. The full definition of the cluster length thus becomes: 

\begin{equation}
\ell(y) = \min\left(\frac{d}{\sqrt{|I(y)|}} , |y - y_\text{w}|\right)
\label{eq:ell_cap}
\end{equation}

\noindent where, $y_\text{w}$ is the position of the nearest wall or layer of change in sign of the shear stress ratio $\mu$. We solve the model defined by Eqs. (\ref{eq:NL_continuum}) and (\ref{eq:ell_cap}) using an implicit finite difference scheme  (\emph{bvp5c} in MATLAB). To be consistent with the boundary conditions used for the Cooperative model, a null gradient in shear rate was prescribed at the boundary.  Results showed on Fig. \ref{fig:sg}(d) and Fig. \ref{fig:vc}(d) shows that this model captures the measured inertial profiles with some discrepancies. The level of discrepancies of this model and of the cooperative model appears to be similar in magnitude. 

Remarkably, while both models lead to similar results in these flow geometries, the profiles of length scale they involved is significantly different. In particular, the cooperative length scale diverges at the location where $\mu=\mu_0$, while the the cluster size defined by (\ref{eq:ell_cap}) always remains smaller than the system size and the nearest distance to a boundary. The fact that the two models provides similar results with such different length scales indicate their low sensitivity to the profile of this parameter.

\begin{figure}[!t]
 \centering
\includegraphics[width=1\columnwidth]{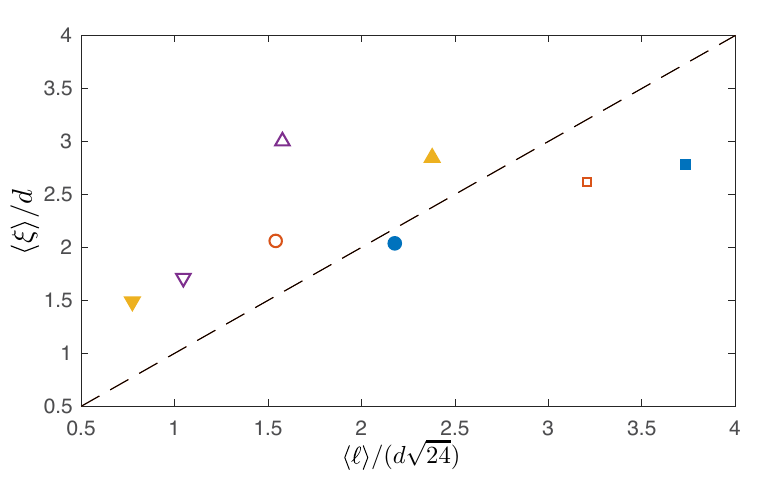}
\caption{\label{fig:mean_ell} 
Average cluster size $\langle \ell \rangle$ from (\ref{eq:ell_cap}) vs the average cooperativity length $\langle \xi \rangle$ from for various DEM simulation configurations shown in table \ref{tab:config}. The dashed line with equation $ \langle \xi \rangle = \langle \ell \rangle/\sqrt{24} $ is presented as a guide for the eye.
}
\end{figure}

Furthermore, figure (\ref{fig:mean_ell}) shows the average cluster size $\langle \ell \rangle =\frac{1}{H}\int^H_{y=0} \ell(y) dy$ and average cooperativity length $\langle \xi \rangle=\frac{1}{H}\int^H_{y=0} \xi(y) dy$ across plane shear and Poiseuille flows. In spite of there difference in profiles, the average cluster size and cooperativity length appear to be similar. This suggests that the two non-local models may be more sensitive to the spatial average of their length scale than to their length scale profile.

\section{Conclusions}

The key result of this paper is that a non-local continuum model (\ref{eq:NL_continuum}) can be derived from a purely kinematic argument, considering the existence of jammed clusters in the flow and the way they spatially redistribute the vorticity.
This derivation indicates that the presence of jammed clusters in the flows can lead to non-local behaviours at the continuum scale.
Clusters were evidenced here for frictional dense granular flows, and there average size was found to follow a power law with the inertial number represented by equation (\ref{eq:xi}). 

More generally, correlated motion of particles \cite{khosropour_convection_1997,radjai_turbulentlike_2002,rognon_thermal_2010}, and avalanche-like rearrangements have been observed in various glassy materials like 
sand \cite{abedi2012vortex}, granular materials \cite{pouliquen_velocity_2004,keys_measurement_2007,henkes2016rigid}, cohesive grains \cite{rognon2008dense}, 
foams \cite{durian1997bubble}, 
suspensions \cite{goyon_spatial_2008,during_length_2014}, colloids \cite{berthier_direct_2005,ballesta_unexpected_2008}, and Lennard-Jones glass \cite{PhysRevE.60.3107,lemaitre2009rate}.
These observations suggest that clusters of particles may exist in many glassy materials. Our derivation indicates that non-local behaviours should then be expected. Furthermore, the link made here between clusters and non-locality creates a promising opportunity to better interpret, analyse and unify the non-local behaviours of glassy materials through a thorough characterisation of cluster size and formation mechanisms for specific materials and flow configurations.

The cluster-based non-local model we derived from the local kinematics is defined by Eqs. (\ref{eq:NL_continuum}) and (\ref{eq:ell_cap}). It only marginally differs from the Cooperative non-local model in that  it does not involve a shear stress or shear to normal stress ratio in the definition of fluidity. With the flow geometries we have considered, we found that this difference did not lead to significant discrepancies between the models, which were both matching relatively well the measured flow profiles. Nonetheless, it is possible that other geometries could reveal some more noticeable discrepancies.

In this work, we have investigated the non-local behaviours by deliberately focusing on micro-kinematic process based on quantifiable cluster sizes.
The fact that this kinematic description could be established does not rule out an alternative interpretation of non-locality in terms of contact forces, length of force chains and diffusion of stress fluctuations through the contact network. 
While a non-local description based on a quantifiable micro-dynamical process would certainly complement our understanding of non-locality, such a description is yet to be established. 

\providecommand*{\mcitethebibliography}{\thebibliography}
\csname @ifundefined\endcsname{endmcitethebibliography}
{\let\endmcitethebibliography\endthebibliography}{}

\end{document}